\documentclass[arxiv]{raa}
\usepackage{natbib}
\usepackage{graphicx,times}

\begin{document}

\title{The evolution of the eccentricity in the eclipsing binary system AS~Camelopardalis}

\author{ V. S. Kozyreva
\inst{1}
\and A. V. Kusakin
\inst{2}
\and A. I. Bogomazov
\inst{1}
}

\institute{M. V. Lomonosov Moscow State University, P. K. Sternberg Astronomical Institute,\\119991, Universitetskij prospect, 13, Moscow, Russia\\{\it a78b@yandex.ru (AIB)}\\
\and
National Space Agency, V. G. Fesenkov Institute of Astronomy, Kazakhstan,\\050020, Observatory 23, Almaty, Kazakhstan\\
}

\abstract{In 2002, 2004, and 2017 we conducted high precision CCD photometry observations of the eclipsing binary system AS~Cam. By the analysis of the light curves from 1967 to 2017 (our data + data from the literature) we obtained photometric elements of the system and found the change of the system's orbital eccentricity by $\Delta e=0.03 \pm 0.01$. This change can indicate that there is a third companion in the system in a highly inclined orbit with respect to the orbital plane of the central binary, and its gravitational influence may cause the discrepancy between the observed and theoretical apsidal motion rates of AS~Cam.
}

\maketitle

\section{Introduction}
\label{sect:intro}

AS Cam is a main-sequence eclipsing binary star (B8V+B9.5V components), its orbital period is $\approx 3.43$ days, the orbital eccentricity is $e\approx 0.17$, the maximum visual brightness is $\approx 8.57^\mathrm{m}$ (Simbad database \footnote{http://simbad.u-strasbg.fr}). It was found in photographic plates by \citet{strohmeier1968}. \citet{hilditch1969,hilditch1972} conducted photoelectric observations of AS Cam, obtained its radial velocity curve and calculated the system's absolute parameters.

\citet{khaliullin1983} discovered the apsidal motion in AS Cam using the {\it WBVR} photometry. The obtained rate of the periastron movement $\dot\omega_\mathrm{obs}=16^{\circ}$ per century was almost 3 times less than the expected theoretical value $\dot\omega_\mathrm{th} = 44^{\circ}$ per century. This discovery was independently cofirmed by \citet{maloney1991}, \citet{wolf1996}. AS Cam became the second (after DI Her, see \citealp{martynov1980}) eclipsing system in which the apsidal motion proved to be much slowly than it was predicted by the theory. To explain the discrepancy between observational estimations and theoretical calculations of the apsidal motion rate in both binaries (DI Her and AS Cam) different authors introduced different hypothesises \citep{shakura1985,moffat1989,claret1997,claret1998}. Most of their hypothesises were already discussed by \citet{maloney1991,claret1997,claret1998}, so we did not describe them here. \citet{zakharov1988}, \citet{khaliullin1991} explained the observed anomalies in the frames of the classical and relativistic mechanics: the gravitational influence of the third companion on the central binary in case of non-coplanar orbits can slow down the apsidal motion. \citet{borkovits2007} significantly improved this idea and enriched it by numerical and analytical computations.

According to \citet{claret2010} there is still no evidence for the existence of the third companion in DI Her, its light equation cannot be found within existing errors of observational data. A leading idea that explains the system's slow apsidal motion is a non-coplanar axial rotation of the stars, \citet{albrecht2009}, \citet{albrecht2011} observed the Rossiter-McLaughlin effect in DI Her and NY Cep. This model requires very high equatorial velocities (up to 300 km s$^{-1}$). In case of AS Cam \citet{kozyreva1999} found an evidence in favour of the existence of the third body. They obtained high precision light curves of AS Cam in 1992-1996, accumulated 10 primary and 13 secondary minima and used them to find variations of times of minima. Based on these data and times of minima from the literature \citet{kozyreva1999} discovered cyclical in-phase variations of primary and secondary times of minima and explained their result by the influence of the third companion, its orbital period was found to be about 805 days, the eccentricity was $\approx 0.5$, the amplitude of the light time effect was $\approx 0.50$ astronomical units.

After the discovery of the light equation in AS Cam in 1999 a lot of new minima times were obtained. So, there is a possibility to re-test the system on the existence of the third companion with higher precision.

\section{Observations}
\label{sect:Obs}

Photometric observations of AS Cam were conducted in 2002, 2004, and 2017 in Tien Shan observatory, Fesenkov Astrophysical Institute (Kazakhstan). In 2002 and 2004 we used the 50 cm AZT-5 telescope with the photomultiplier tube (PMT) model 79 and {\it V} filter, the comparison star was HD 34463. In 2017 we obtained new CCD observations in {\it B}, {\it V}, and {\it R} filters using the Zeiss-1000 telescope equipped with the Apogee U900 CCD camera. For the latest set of observations we used TYC 4347-452-1 (comparison star) and TYC 4347-682-1 (control star) as reference stars. Usual exposure
times were about 10 seconds.

To process raw CCD data we used the MAXIM-5 program. The aperture was constant during one night, its differences from night to night was not significant. Maximum errors for a single exposure were in the range $0.003^m-0.006^m$ for the different nights. Reference stars were assumed to be constant during observations. For the values in Tables \ref{ourmoments} and \ref{parevolution} we used only the best light curves in {\it V} filter with standard deviation less than $0.007^m$. Standard dark and flat
field corrections were made. In order to obtain the maximum possible precision of times of minima we used only full light curves between their maxima, our values obtained for 2002, 2004, and 2017 observations are shown in Table \ref{ourmoments}.

\section{Light equation}
\label{sect:light}

A computer code was used to find orbital elements and system parameters. A description of the method used in the code to solve light curves can be found in Section ``Algorithms and models'' in the paper by \citet{kozyreva2001}, a similar model was described by \citet{khaliullin1984}. The code seeks for the photometric parameters and orbital elements using a simple model of two spherical stars (with a linear limb darkening law) that move around a common center of masses in elliptical orbits. The parameters are: the radii of the primary and secondary components $r_{1, 2}$, the limb darkening coefficients for the components $u_{1, 2}$, the luminosities of components in fractions of the system's total luminosity $L_{1, 2}$, the inclination of the orbit of the binary with respect to the plane of the sky $i$, the orbital eccentricity $e$, the longitude of periastron of the primary star's orbit $\omega$, the epoch of the primary minimum corresponding to the epoch of the observations analysed, the system's third light parameter $L_3$. AS Cam is a pair of definitely detached stars, therefore the model of two spherical stars with limb darkening is quite reasonable. The solution was accepted as adequate only when ``observed minus calculated'' value was the lowest and it had no systematic deviations within the minima. The values of parameters were found in a free search excluding the limb darkening coefficients. The values of limb darkening coefficients ($0.46\pm 0.03$ and $0.31\pm 0.05$ for the primary and secondary respectively) were taken from the paper by \citet{khaliullin1983}, they were computed using the high quality light curves of AS Cam obtained in 1981. The values of other parameters are slightly different from year to year being within the error bars of their values in Table III in \citet{khaliullin1983}, except $i$ and $e$ (see Table \ref{parevolution}).

We took into account all times of minima from the Table \ref{ourmoments} and from the B.R.N.O. database\footnote{http://var2.astro.cz/ocgate/index.php?lang=en}. The AS Cam apsidal motion is slow, its period is about 2400 years. During 55 years of observations this rate yields only $\approx 8^{\circ}$ of the total cycle ($360^{\circ}$), therefore it is possible to use linear approximations instead of sinusoidal changes of times of minima to investigate the possible light equation in the system. We obtained following ephemerides for the primary and secondary minima with the same orbital period of the central binary:

\begin{eqnarray}
C_1(\textrm{Min}\ \textrm{I}) & = & \textrm{HJD}2440125.60300 + 3.43096730\times E, \\
C_1(\textrm{Min}\ \textrm{II}) & = & \textrm{HJD}2440123.67395 + 3.43096730\times E.
\end{eqnarray}

\noindent where $E$ is the number of orbital cycles since the initial epoch, HJD is the Heliocentric Julian Date of the initial epoch.

\begin{figure}
\centering
\includegraphics[width=\textwidth, angle=0]{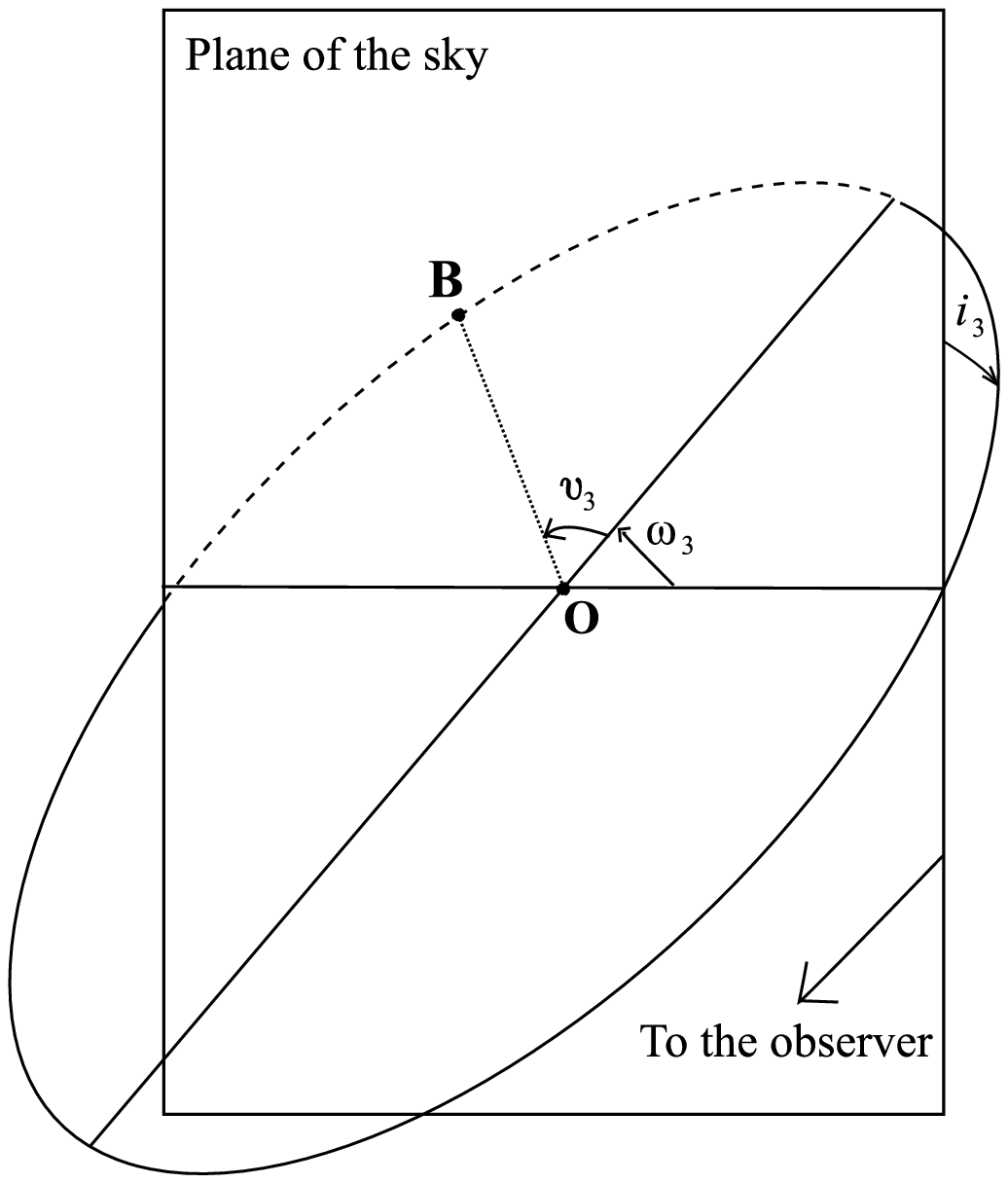}
\caption{A reference frame for orbital elements of the third body. Here "O" is the center of masses of the triple system, "B" is the third body within the orbit around "O", $i_3$ is the inclination of the orbit, $\omega_3$ is the periastron longitude, $v_3$ is the true anomaly.}
\label{figure}
\end{figure}

The presence of the third body in an eclipsing binary system can observationally appear as the periodical variations of its times of minima in comparison with the system's linear ephemerides. Such variations arise from the motion of the center of masses of the binary star around the center of masses of the triple system. The amplitude of variations for primary minima is given by a light equation:

\begin{equation}
(O-C)_1 = \frac{a_3\sin{i_3}}{c}(1-e_3\cos{E_3})\sin(v_3+\omega_3),
\label{lighteq}
\end{equation}

\noindent where $v_3$ is the true anomaly of the third companion's orbit, $E_3$ is its eccentric anomaly, $a_3$ is the semi-major axis of the third companion's orbit, $i_3$ is the inclination of this orbit with respect to the plane of the sky, $e_3$ is its eccentricity, and $\omega_3$ is the pericentric longtitude. Figure \ref{figure} shows the reference frame for the third body's orbit. The value of $E_3$ is connected with other elements as follows:

$$
\frac{2\pi}{P_3}(t-T_3)=(E_3-e_3\sin E_3),
$$
\noindent where  $T_3$ is the time of the periastron passage by the third body, $t$ is the time, $P_3$ is the orbital period of the third body.

$(O-C)_1$ experiences periodical variations in the time scale $\approx 2$ years (see Figures \ref{figure1} and \ref{figure2}), these variations have the same phase for the primary and secondary minima. It is essential that both sets of observations (1968-1973 in Figure \ref{figure1} and 1980-2017 in Figure \ref{figure2}) can be described by the same light equation curve. The estimations of the parameters in Equation (\ref{lighteq}) that we obtained (we applied the least-squares method for the $(O-C)_1$) are listed in Table \ref{lighteqparams}, and the ephemerides for the third companion are following:

\begin{equation}
\label{eph3}
\textrm{Min}\ \textrm{III} = \textrm{HJD} 2444265.036 + 805.9\times E_1.
\end{equation}

\noindent where $E_1$ is the number of orbital cycles of the third body since the initial epoch.

The mass function

\begin{equation}
f(m) = \frac{(M_3\sin i_3)^3}{(M_1+M_2+M_3)^2} = \frac{(a_3\sin i_3)^3}{P_3^2}.
\label{mf}
\end{equation}         
                
\noindent gives (after the subtraction of the parameters of the light equation) the lower limit of the third body's mass $M_3 \sin i_3 \approx 1.1  M_{\odot}$.

The light equation method allows to estimate only the lower limit of the third body's mass, and the real value depends also on the angle between the plane of the sky and the orbital plane of third companion. If the angle $i_3\leq 30^{\circ}$, the mass of the third body should be $\geq 2.2 M_{\odot}$ (this value is comparable with the secondary star of the central tight binary). Our photometric solutions for the light curve can be plausible only if the luminosity of the third companion is no more than 3.5\% of the total luminosity of the system. If this body is a non-degenerate main-sequence companion, its mass is less than $1.5 M_{\odot}$ and $i_3\geq 43^{\circ}$. Also the spectral lines of the third companion were not found, therefore the mass of a hypothetical main-sequence companion has an upper limit. In general, the suggested body can be a compact remnant (a white dwarf or even a neutron star) or can be a very close binary star like YY Gem.

\section{AS Cam apsidal motion rate re-estimation}

The difference between periods of primary and secondary minima is the indication of the apsidal motion rate (see Figure \ref{figure3}). Using the least square method we computed following ephemerides:

\begin{eqnarray}
C_2(\textrm{Min}\ \textrm{I}) & = & \textrm{HJD}2444939.24519 + 3.43096365\times E, \\
C_2(\textrm{Min}\ \textrm{II}) & = & \textrm{HJD}2444937.32569 + 3.43097095\times E.
\end{eqnarray}

These equations correspond to the rate of the apsidal motion $\dot{\omega}_\mathrm{obs}= 15.5^{\circ}\pm 1.5^{\circ}$ per century. The theoretical apsidal motion rate for AS Cam was found to be $\dot{\omega}_\mathrm{th} = \dot{\omega}_\mathrm{cl} + \dot{\omega}_\mathrm{rel} = 44^{\circ}$ per century \citep{maloney1991}.

The parameters obtained in this study coincide with the parameters of the light equation and of the apsidal motion rate calculated by \citet{kozyreva1999} within the error bars, despite the fact that the results in 1999 were found from much less observational data.

\section{AS Cam slow apsidal motion possible causes}

The influence of the third companion on the apsidal motion rate in general was studied by \citet{khaliullin1991}, \citet{khodykin1997}, \citet{khodykin2004}. The low apsidal motion rate in DI Her and in AS Cam probably could be explained by the perturbations cause by the third body. This idea was improved by \citet{borkovits2007}, they conducted theoretical investigations of configurations of the central binary's orbit and the third companion's orbit using AS Cam as an example. \citet{borkovits2007} considered four variants of the mutual disposition of both orbits: the angle between them equals $0.8^{\circ}$, $20^{\circ}$, $60^{\circ}$, and $90^{\circ}$. The results were shown in figures 6-9 by \citet{borkovits2007} for the fast (in a time scales of about $\leq 100$ years) and for the slow (in a time scales of about 3000 years) evolution of orbital parameters.
       
It is possible to compare the values of parameters obtained in past (since 1981 for our data) and modern (2017) observations for the short time scale, a large amount of light curves within primary and secondary minima was accumulated. We took light curves with the precision better than 1\% within light curve minima in our data, and the light curve published by \citet{hilditch1972}. It allowed to compute orbital elements, e. g. \citep{kozyreva2001}. We used our light curves of AS Cam obtained in 1981, 1992,1993,1994,1995,1996,2002, 2017, and light curves obtained by \citet{hilditch1972} in 1967 and 1968. The dataset covered 50 years of observations and allowed to trace the change of the eccentricity and the inclination of the central binary's orbit (see Table \ref{parevolution}) and to compare the observed changes with the theory of \citet{borkovits2007}. The eccentricity change is $\approx 0.03 \pm 0.01$ during 50 years of observations, the inclination change during the same period of time is $\approx 1 \pm 0.5^{\circ}$. Our value of the eccentricity change is higher than its theoretical value for mutual inclination of the orbits $i'=0$ and $i'=20^{\circ}$, it is almost the same as the theoretical value calculated for $i'=60^{\circ}$, see \citet{borkovits2007}, Figure 2. Their paper contains only four values of mutual inclination of the orbits, therefore it is possible that $i'$ is less than $60^{\circ}$. For the case of $i'=60^{\circ}$ \citet{borkovits2007} showed that the dependence of the difference between primary and secondary minima is more complicated than a simple difference between two sinusoidal functions shifted relative to each other by $180^{\circ}$. The dependence between times of minima is complicated and has non-uniform character, so the apsidal motion rate can change in different periods of time, see Figure 3 by \citet{borkovits2007}.  Such periods with different apsidal motion rates can have duration up to several centures, and now we can see AS Cam in a form of slow apsidal motion.

There is another explanation of the slow apsidal motion rate, it is the spin-orbit misalignment \citep{albrecht2009,albrecht2011}. \citet{pavlovski2011} obtained spectral lines of AS Cam and found that the lines are narrow in comparison with the expected line width in supposition of synchronous rotation of the stars. They concluded that the axial and orbital rotation are not aligned. \citet{shakura1985} found that the apsidal line can undegro retrograde motion if the
rotation axises of the stars are not aligned to the orbital axis, with the largest effect when the axises are perpendicular. If the stars were rotating at three times the synchronous rate around axises that lie almost in the orbital plane, and if one takes into account the projected rotational velocities, the stars' rotational axes should be tilted by $82^{\circ}$ and $87^{\circ}$  with respect to the orbital axis \citep{pavlovski2011}. If the orbit of that body is highly inclined with respect to the orbit of the inner two stars, then the Lidov-Kozai mechanism \citep{lidov1962,kozai1962} may be invoked to explain the spin-orbit misalignment. 

Probably, both mechanisms (the influence of the third body and the spin-orbit misalignment) of the deceleration of the apsidal motion rate are plausible for the AS Cam binary.

\section{Conclusions}
\label{sect:conclusion}

During the 50 years time interval the orbit of the close binary star AS Cam changed its eccentricity by $\approx 0.03$. This fact indicates that the system possesses the third body in a highly inclined orbit with respect to the central binary's orbit. Such configuration can induce the precession of rotational axises of the stars in the binary, and trigger of their spin-orbit misalignment. Both mechanisms can slow down the apsidal motion rate in comparison the system without a third companion, and if the rotational axises were aligned with the orbital axis.

The unresolved question is how the third body with the mass comparable to the masses of the stars in central binary does not show itself neither in spectra of the stars, nor in photometric elements obtained as a solution for the light curves. This body can be a degenerate star or a very close binary system like YY Gem consisting of low luminosity main-sequence stars.

\begin{acknowledgements}
This study was supported by by the funding program BR05236322 of the Ministry of Education and Science RK and by a grant for Leading Scientific Schools NSh-9670.2016.2 (Russia). We are grateful to A. I. Zakharov for the software and to M. A. Krugov for his efforts to establish the collaboration with Tien Shan observatory. We thank the anonymous referee for useful comments.
\end{acknowledgements}

\clearpage

\clearpage

\begin{figure}
\centering
\includegraphics[width=\textwidth, angle=0]{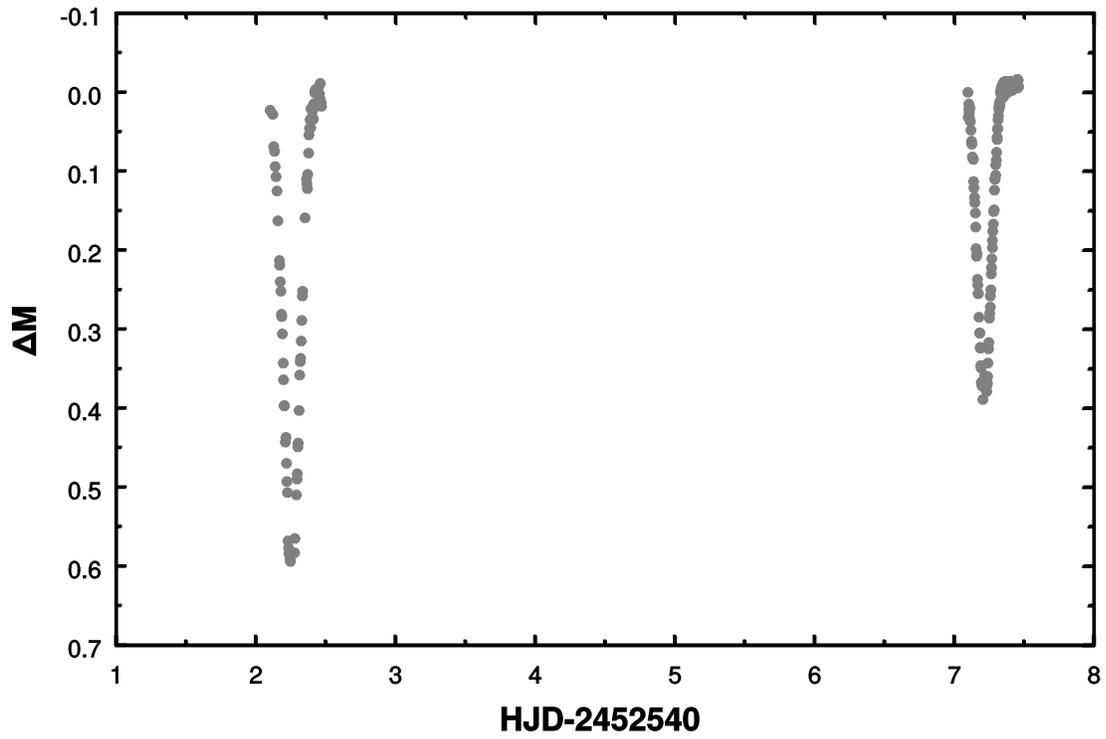}
\caption{A sample light curve of AS~Cam. $\Delta M$ is the change of its magnitude, HJD is the Heliocentric Julian Date.}
\label{figure0}
\end{figure}

\clearpage

\begin{figure}
\centering
\includegraphics[width=\textwidth, angle=0]{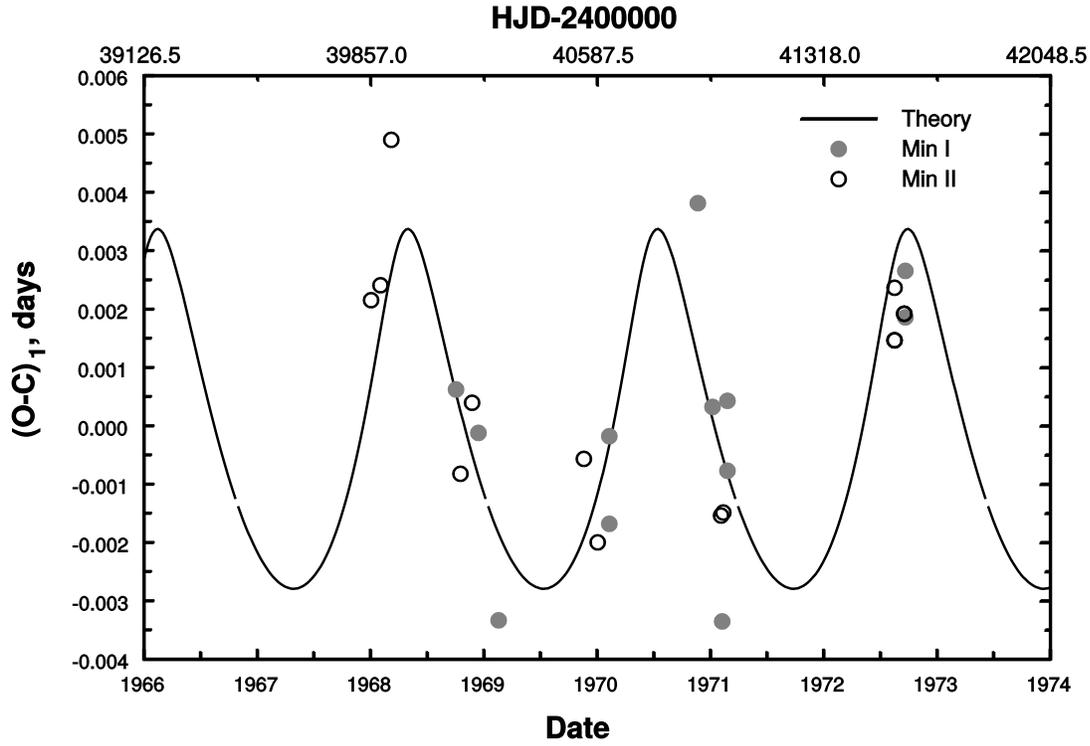}
\caption{The light equation curve (``Theory'' in Figure) for $(O-C)_1$ calculated as the difference between observed times of minima (Min I and Min II) and values computed using Equations 1-2 for observations in 1967-1973. See Table \ref{lighteqparams} for parameters of the third body's orbit. HJD is the Heliocentric Julian Date, the bottom horizontal axis depicts the time in years.}
\label{figure1}
\end{figure}

\clearpage

\begin{figure}
\centering
\includegraphics[width=\textwidth, angle=0]{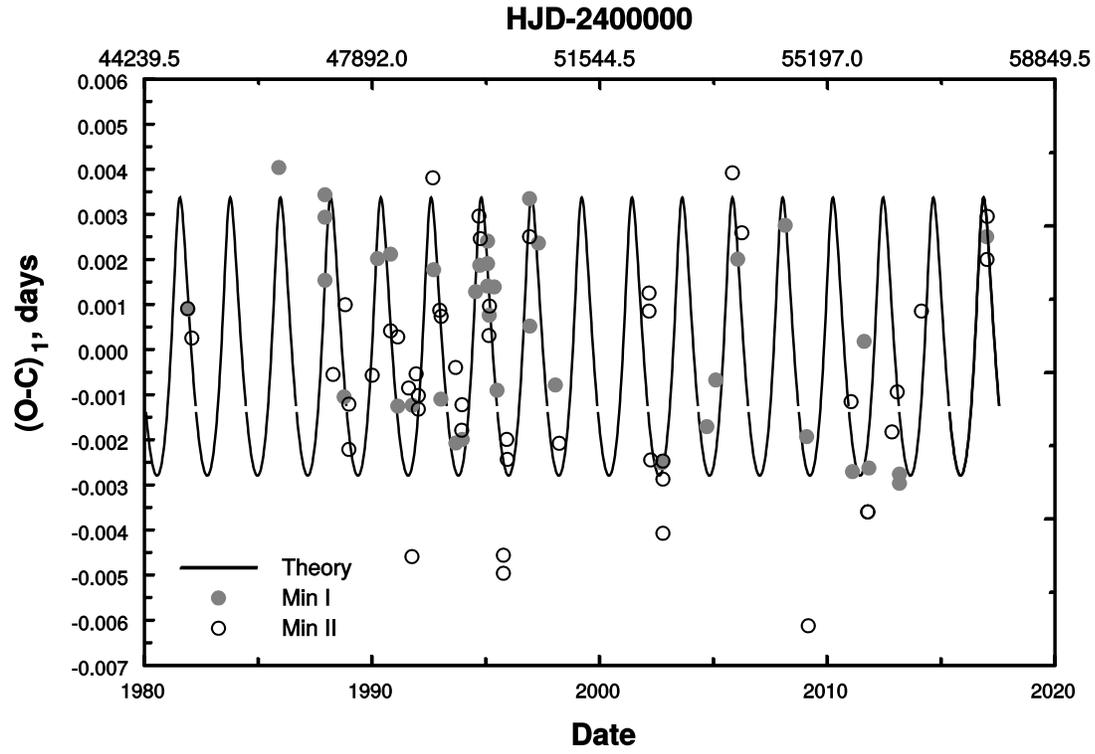}
\caption{The same as Figure \ref{figure1} for observations in 1981-2017.}
\label{figure2}
\end{figure}

\clearpage

\begin{figure}
\centering
\includegraphics[width=\textwidth, angle=0]{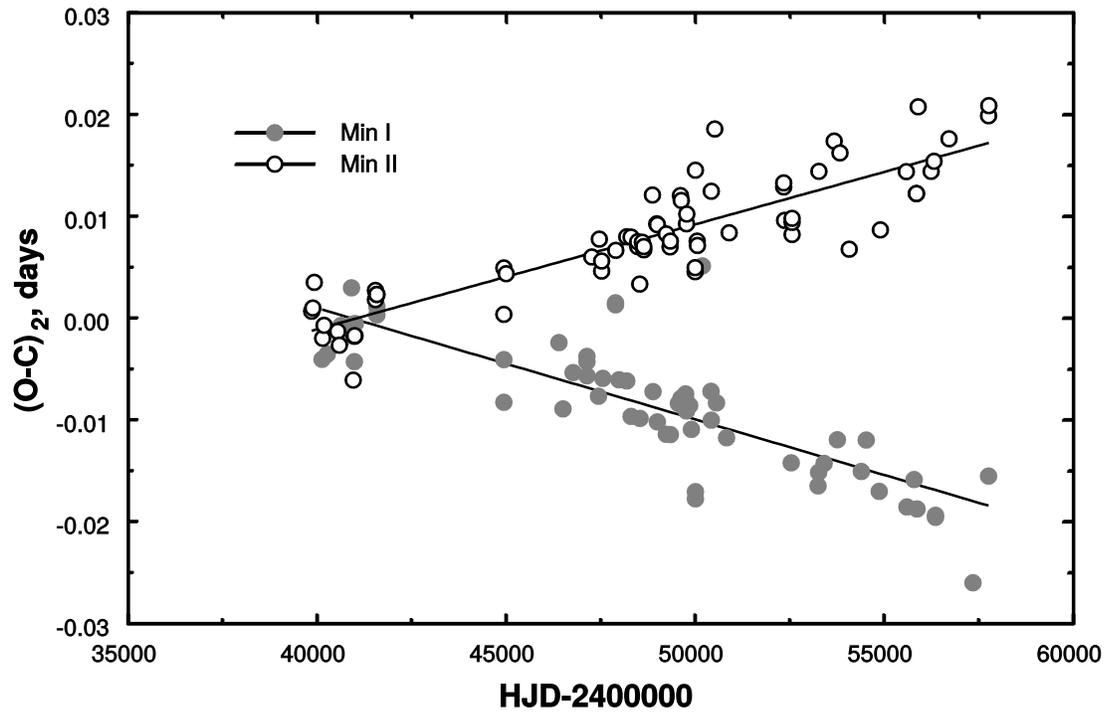}
\caption{The discrepancy between primary and secondary minima. It illustrates the apsidal motion in the system. $(O-C)_2$ calculated as the difference between observed times of minima (Min I and Min II) and values computed using Equations 6 and 7. HJD is the Heliocentric Julian Date.} 
\label{figure3}
\end{figure}

\clearpage

\begin{table}
\begin{center}
\caption{AS Cam Times of Minima Obtained in This Study. See Equations 1 and 2 for $(O-C)_1$ and Equations 6 and 7 for $(O-C)_2$. HJD is the Heliocentric Julian Date, the ``Min'' column depicts the type the minimum (primary or secondary).}
\label{ourmoments}
\begin{tabular}{cccc}
\hline\noalign{\smallskip}
HJD$-2400000$ & Min & $(O-C)_1$, days & $(O-C)_2$, days \\
\hline\noalign{\smallskip}
52542.2593 & I & -0.01436 & -0.00145 \\
52547.2206 & II & 0.01406 & 0.00134 \\
53252.4673 & I & -0.01659 & -0.00296 \\
53266.1925 & I & -0.01526 & -0.00161 \\
53271.1555 & II & 0.01486 & 0.00137 \\
57757.3284 & I & -0.01554 & 0.00289 \\
57762.2973 & II & 0.02046 & 0.00220 \\
57769.1602 & II & 0.02143 & 0.00316 \\
\hline\noalign{\smallskip}
\end{tabular}
\end{center}
\end{table}

\begin{table}
\begin{center}
\caption{The values of the parameters in Equation (\ref{lighteq}). See text for details.}
\label{lighteqparams}
\begin{tabular}{cc}
\hline\noalign{\smallskip}
Parameter & Value \\
\hline\noalign{\smallskip}
$\frac{a_3\sin i_3}{c}$ & $4.5\pm 0.3$ min \\
$e_3$ & $0.42\pm 0.05$ \\
$P_3$ & $805.9\pm 1.5$ days \\
$\omega_3$ & $69\pm 2$, $^{\circ}$ \\
\hline\noalign{\smallskip}
\end{tabular}
\end{center}
\end{table}

\begin{table}
\begin{center}
\caption{The eccentricity $e$ and inclination $i$ of the orbit of AS Cam calculated using the full dataset.}
\label{parevolution}
\begin{tabular}{ccc}
\hline\noalign{\smallskip}
Year & $e$ & $i$ \\
\hline\noalign{\smallskip}
1967-1968 & $0.147\pm 0.010$ & $88.3\pm 0.4^{\circ}$ \\
1981 & $0.167\pm 0.008$ & $88.6\pm 0.4^{\circ}$ \\
1992 & $0.161\pm 0.015$ & $88.8\pm 0.5^{\circ}$ \\
1993 & $0.166\pm 0.010$ & $88.6\pm 0.6^{\circ}$ \\
1994 & $0.166\pm 0.013$ & $88.9\pm 0.5^{\circ}$ \\
1995 & $0.170\pm 0.010$ & $89.0\pm 0.5^{\circ}$ \\
1996 & $0.164\pm 0.016$ & $88.8\pm 0.5^{\circ}$ \\
2002 & $0.164\pm 0.010$ & $89.3\pm 0.5^{\circ}$ \\
2017 & $0.178\pm 0.008$ & $89.5\pm 0.5^{\circ}$ \\
\hline\noalign{\smallskip}
\end{tabular}
\end{center}
\end{table}

\label{lastpage}

\end{document}